\documentclass[aps,twocolumn,prl,10pt,superscriptaddress,nobibnotes,showpacs]{revtex4}
\usepackage{amsfonts}
\usepackage{amsmath}
\usepackage{amssymb}
\usepackage{graphicx}

\providecommand{\U}[1]{\protect\rule{.1in}{.1in}}

\begin{document}

\title{Localized whistlers in magnetized spin quantum plasmas}
\author{A. P. Misra}
\email{apmisra@visva-bharati.ac.in}
\altaffiliation{Permanent address: Department of Mathematics, Siksha Bhavana, Visva-Bharati
University, Santiniketan-731 235, India.}
\affiliation{Department of Physics, Ume{\aa } University, SE--901 87 Ume{\aa }, Sweden.}
\author{G. Brodin}
\email{gert.brodin@physics.umu.se}
\affiliation{Department of Physics, Ume{\aa } University, SE--901 87 Ume{\aa }, Sweden.}
\author{M. Marklund}
\email{mattias.marklund@physics.umu.se}
\affiliation{Department of Physics, Ume{\aa } University, SE--901 87 Ume{\aa }, Sweden.}
\author{P. K. Shukla}
\email{ps@tp4.rub.de; profshukla@yahoo.com}
\affiliation{Department of Physics, Ume{\aa } University, SE--901 87 Ume{\aa }, Sweden.}
\affiliation{RUB International Chair, International Centre for Advanced Studies in
Physical Sciences, Faculty of Physics \& Astronomy, Ruhr University Bochum,
D-44780 Bochum, Germany.}

\begin{abstract}
The nonlinear propagation of electromagnetic (EM) electron-cyclotron waves
(whistlers) along an external magnetic field, and their modulation by
electrostatic small but finite amplitude ion-acoustic density perturbations
are investigated in a uniform quantum plasma with intrinsic spin of
electrons. The effects of the quantum force associated with the Bohm
potential and the combined effects of the classical as well as the
spin-induced ponderomotive forces (CPF and SPF respectively) are taken into
consideration. The latter modify the local plasma density in a
self-consistent manner. The coupled modes of wave propagation is shown to be
governed by a modified set of nonlinear Schr\"{o}dinger-Boussinesq-like
equations which admit exact solutions in form of stationary localized
envelopes. Numerical simulation reveals the existence of large-scale density
fluctuations that are self-consistently created by the localized whistlers
in a strongly magnetized high density plasma. The conditions for the
modulational instability (MI) and the value of its growth rate are obtained.
Possible applications of our results, e.g., in strongly magnetized dense
plasmas and in the next generation laser-solid density plasma interaction
experiments are discussed.
\end{abstract}

\date{23 Oct., 2010}
\maketitle

\section{I. Introduction}

Having been discovered more than a century ago \cite{WhistlerDiscovery},
whistler waves become one of the most important waves in plasmas. Such waves
(also known as helicons in solid state plasmas) are low-frequency (lf) (in
comparison with the electron-cyclotron frequency, $\omega_{c}$)\ right-hand
circularly polarized (RCP) electromagnetic (EM) waves guided almost along
the external magnetic field in dense plasmas. Because of the increase of
their group velocity with the frequency, $\omega<\omega_{c}/2$ (see, e.g.
Ref. \cite{JPPMisra}), the lf waves arrive later giving rise a whistling
down-effect observed at ground level. Stenzel in his classic paper \cite%
{Stenzel} demonstrated experimentally the creation of magnetic field-aligned
density perturbations excited by the ponderomotive force exerted by the EM
whistlers.

Whistler waves are important not only in space plasmas due to wave-particle
interactions, but also in laboratory plasmas as helicons for efficient
plasma production as well as in dense astrophysical environments \cite%
{WhistlerTheory1,WhistlerTheory2,MI,WhistlerTheory3(nearsonic),WhistlerTheory4,WhistlerTheory5,WhistlerParametricInstability}%
. On the other hand, large amplitude whistlers propagating in a magnetized
plasma can initiate a great variety of nonlinear effects, e.g., three-wave
interactions, parametric instabilities \cite{WhistlerParametricInstability},
modulational instability and the subsequent soliton formation \cite%
{WhistlerTheory1,WhistlerTheory2,MI}. The latter which, in turn, causes
local electron density enhancement or depletion in plasmas, are considered \
as a basis for understanding laser energy deposition in pellets \cite{Pellet}%
, pulsar radiation interaction with the ambient magnetosphere \cite%
{AmbientMagnetosphere}, whistler wave propagation in solar winds \cite%
{SolarWind} etc. \ Recent laboratory experiment \cite{WhistlerExperiment2}
and observations from the Freja satellite \cite{FrejaSatellite} show the
clear evidence for the formation of whistler envelope solitons accompanied
by plasma density cavities. Moreover, electrons in Van Allen radiation belts
can be accelerated \ to MeV energies within a short period by large
amplitude whistlers \cite{Cattell}. The latter have recently been observed
by the Cluster spacecraft \cite{ClusterSpacecraft}, the STEREOS \cite%
{Cattell} and the THEMIS \cite{THEMIS}. Furthermore, laboratory experiments 
\cite{Spheromak1} and theoretical confirmation \cite{Spheromak2} have
demonstrated the existence of propagating whistler spheromaks with fields
exceeding the ambient magnetic field. Whistlers also contribute to fast
magnetic reconnection and plasma dynamics in two-beam laser-solid density
plasma interaction experiments \cite{Laser-solidInteraction}.

Recently, there has been a notably growing interest in investigating various
quantum plasma effects in view of some experimental progresses in nanoscale
plasmas \cite{QuantumWell}, ultracold plasmas \cite{Ultracold}, spintronics 
\cite{Spintronics} and plasmonics \cite{Plasmonics}. On the other hand,
superdense quantum plasmas are omnipresent in compact astrophysical objects,
e.g., the interior of massive white dwarfs, interior of Jupitors, magnetars
etc. \cite{CompactObjects1,CompactObjects2,CompactObjects3}, as well as in
the next generation intense laser-solid density plasma interaction
experiments \cite{LaserInteraction1,LaserInteraction2,LaserInteraction3}. In
dense plasmas, degenerate electrons follow Fermi-Dirac pressure law, and
there are\ typically quantum force associated with the Bohm de Broglie
potential, which produce wave dispersion at nanoscales \cite%
{Quantum1,Quantum2,Quantum3}. Furthermore, the effects of the electron spin
manifests itself in terms of a magnetic dipole force, as well spin
precession, which can be exploited by transforming the Pauli equation to
fluid-like variables \cite{Spin1,Spin3}. More elaborate kinetic models has
also been developed \cite{Spin-kinetic,Spin2}. Hence the dynamics of
electrons in Fermi degenerate plasmas will be affected not only by the
Lorentz force, but also by the effects of quantum statistical pressure, the
Bohm force as well as the effects due to intrinsic spin of electrons. We
ought to mention that in a dense magnetized plasma there also exist spin
waves, which can be excited by intense neutrino fluxes. Thus, nonlinear
theories of EM waves, in particular whistlers in magnetized dense plasmas
need to be developed in its own right accounting for all these quantum
effects. Recently, the theory of the ponderomotive force in plasmas has been
extended to account for the contribution from the intrinsic spin of
electrons \cite{SpinPonderomotive}. It has been demonstrated that an EM
pulse can induce a spin-polarized plasma by this spin-ponderomotive force
(SPF). Such force could also play an important role in the propagation of lf
EM waves, e.g., whistlers, Alfv\'{e}n waves.

Our objective here is to present \ a theoretical study of modulated whistler
wave packets interacting nonlinearly with background lf density
perturbations that are reinforced by the classical ponderomotive force (CPF) 
\cite{WhistlerTheory2} as well as the SPF \cite{SpinPonderomotive}. The role
of the ion motion as well as the dispersive effects due to charge separation
and the electron tunneling are also taken into account. We will include the
field aligned velocity perturbation (free electron streaming) associated
with the lf motion, and in addition, generalize the related classical
results that exist in the literature (see, e.g., Refs. \cite%
{WhistlerTheory1,WhistlerTheory2}). The obtained results could be useful for
understanding the propagation of localized EM whistlers which may emanate in
the interior of magnetized white dwarfs, magnetars as well as in the next
generation intense laser-solid density plasma experiments.

\section{II. Nonlinear evolution equations}

%%\tableofcontents
Let us consider the propagation of nonlinearly coupled EM whistlers and
ion-acoustic (IA) density perturbations along a constant magnetic field $%
\mathbf{B}=B_{0}\hat{z}$ in a quantum electron-ion plasma where any
equilibrium drift velocity is zero. In the modulational representation, the
high-frequency (hf) EM wave field for the RCP whistlers is given by $\mathbf{%
E}=\left( \hat{x}-i\hat{y}\right) E(z,t)\exp (ikz-i\omega t)+$c.c., where $%
E(z,t)$ is the slowly varying (both in space and time) envelope of the
whistler wave electric field and c.c. stands for the complex conjugate.
Also, $\omega $ $(k)$ represents the whistler wave frequency (number). The
basic equations for the evolution of nonlinear whistlers then read \cite%
{Spin1,SpinPonderomotive,SpinEvolutionEqn}.

\begin{equation}
\frac{\partial n_{e}}{\partial t}+\nabla .\left( n_{e}\mathbf{v}_{e}\right)
=0  \label{b1}
\end{equation}%
\begin{align}
&\left( \frac{\partial }{\partial t}+\mathbf{v}_{e}.\nabla \right) \mathbf{v}%
_{e} =-\frac{e}{m_{e}}\left( \mathbf{E}+\mathbf{v}_{e}\times \mathbf{B}%
\right) -\frac{\nabla P_{e}}{m_{e}n_{e}}  \notag \\
&+\frac{\hbar ^{2}}{2m_{e}^{2}}\nabla \left( \frac{\nabla ^{2}\sqrt{n_{e}}}%
{\sqrt{n_{e}}}\right)+\frac{2\mu }{m_{e}\hbar }\mathbf{S}.\nabla B,  \label{b2}
\end{align}%
\begin{equation}
\left( \frac{\partial }{\partial t}+\mathbf{v}_{e}.\nabla \right) \mathbf{S}%
=-\frac{2\mu }{\hbar }\left( \mathbf{B}\times \mathbf{S}\right),  \label{b3}
\end{equation}%
where $n_{e},$ $m_{e},$ $\mathbf{v}_{e}$ denote the number density, mass and
velocity of electrons respectively, $\mathbf{B}$ is the magnetic field and $%
P_{e}$ is the electron thermal pressure. Also, $\mathbf{S}$ is the spin
angular momentum with its absolute value $\left\vert \mathbf{S}\right\vert
=\left\vert S_{0}\right\vert \equiv \hbar /2;$ $\mu =-\left( g/2\right) \mu
_{B},$where $g\approx 2.0023193$ is the electron $g$-factor and $\mu
_{B}\equiv e\hbar /2m_{e}$ is the Bohr magneton. The equations (\ref{b1})-(%
\ref{b3}) are then closed by the following Maxwell equations with $\nabla .%
\mathbf{B}=0$.%
\begin{equation}
\nabla \times \mathbf{E}=-\frac{\partial \mathbf{B}}{\partial t},  \label{b4}
\end{equation}%
\begin{equation}
\nabla \times \mathbf{B}=\mu _{0}\left( \varepsilon _{0}\frac{\partial 
\mathbf{E}}{\partial t}\mathbf{-}en_{e}\mathbf{v}_{e}\mathbf{+}\frac{2\mu }{%
\hbar }\nabla \times n_{e}\mathbf{S}\right) .  \label{b5}
\end{equation}%
The equations (\ref{b1})-(\ref{b3}) represent the nonrelativistic evolution
of spin$-1/2$ electrons, and are applicable even when different states with
spin-up and spin-down (relative to the magnetic field) can be well
represented by a macroscopic average. This may, however, occur in the
regimes of very strong magnetic fields (or a very low temperature regimes),
where generally the electrons occupy the lowest energy spin states. On the
other hand, for a time-scale longer than the spin-flip frequency, the
macroscopic spin state is well-described by the thermodynamic equilibrium
spin configuration, and in this case the above model can still be applied.
However, such case in which the macroscopic spin state will be attenuated by
a factor decreasing the effective value of $\left\vert \mathbf{S}\right\vert 
$ below $\hbar /2$, will not be considered further in the present work . As
a consequence, our studies will be focused on the regime of strong magnetic
fields and high density plasmas.

Taking the curl of Eq. (\ref{b2}) and using Eqs. (\ref{b3})-(\ref{b5}) we
readily obtain the following evolution equation for whistlers. %
%\begin{equation}
\begin{widetext}
\begin{align}
0& =\frac{e}{m_{e}}\frac{\partial \mathbf{B}}{\partial t}+\frac{\varepsilon
_{0}}{en_{e}}\frac{\partial }{\partial t}\left( \frac{\partial ^{2}\mathbf{B}%
}{\partial t^{2}}+\frac{1}{n_{e}}\nabla n_{e}\times \frac{\partial \mathbf{E}%
}{\partial t}\right) -v_{ez}\nabla \times \frac{\partial \mathbf{v}_{e}}{%
\partial z}+\frac{1}{e\mu _{0}}\frac{\partial }{\partial t}\left[ \frac{1}{%
n_{e}}\nabla \times \left( \nabla \times \mathbf{B}\right) \right]   \notag
\\
& -\frac{2\mu }{e\hbar }\frac{\partial }{\partial t}\left[ \frac{1}{n_{e}}%
\nabla \times \left( \nabla \times n_{e}\mathbf{S}\right) \right] +\frac{1}{%
m_{e}\mu _{0}n_{e}}\nabla \times \left[ \left( \nabla \times \mathbf{B}%
\right) \times \mathbf{B}\right] +\frac{2\mu }{m_{e}\hbar }\nabla \times
\left( S^{a}\nabla B_{a}\right)   \notag \\
& -\frac{\varepsilon _{0}}{m_{e}n_{e}}\nabla \times \left( \frac{\partial
\mathbf{E}}{\partial t}\times \mathbf{B}\right) -\frac{2\mu }{m_{e}\hbar
n_{e}}\nabla \times \left[ \left( \nabla \times n_{e}\mathbf{S}\right)
\times \mathbf{B}\right].  \label{ev}
\end{align}%
\end{widetext}In the linear theory, the whistler frequency $\omega $ and the
wave number $k$ are related by the following linear dispersion relation in
the nonrelativistic limit (see for details, Ref. \cite{JPPMisra}).

\begin{equation}
n_{R}^{2}\left( 1+\frac{\omega _{\mu }}{\omega -\omega _{g}}\right) =1-\frac{%
\omega _{pe}^{2}}{\omega \left( \omega -\omega _{c}\right) },  \label{1}
\end{equation}%
where $n_{R}\equiv ck/\omega $ is the refractive index, $\omega _{\mu
}=g^{2}\left\vert S_{0}\right\vert /4m_{e}\lambda _{e}^{2}$ is the frequency
due to the plasma magnetization current and $\lambda _{e}\equiv c/\omega
_{pe}$ is the electron skin depth with $\omega _{pe(i)}\equiv \sqrt{%
n_{0}e^{2}/\varepsilon _{0}m_{e(i)}}$ denoting the electron (ion) plasma
frequency. Also, $\omega _{c}=eB_{0}/m_{e}\ $is the electron-cyclotron
frequency and $\omega _{g}=(g/2)\omega _{c}$ \ is the electron
spin-precession frequency.

The nonlinear dynamics of whistler wave envelopes under the modulation of
electron density perturbations associated with the lf IA fluctuations and of
the nonlinear frequency-shift caused by the magnetic field aligned free
streaming of electrons with flow speed $v_{ez},$ can be described by the
following nonlinear Schr\"{o}dinger (NLS)-like equation which is obtained
from the EM wave equation (\ref{ev}) as 
\begin{equation}
i\left( \frac{\partial E}{\partial t}+v_{g}\frac{\partial E}{\partial z}%
\right) +\frac{v_{g}^{\prime }}{2}\frac{\partial ^{2}E}{\partial z^{2}}%
-\Delta E=0,  \label{2}
\end{equation}%
where $E\equiv E_{x}-iE_{y},$ and the group speed, $v_{g}\equiv d\omega /dk$
[see Eq.$\ $(11) in Ref. \cite{JPPMisra}] and the group dispersion, $%
v_{g}^{\prime }\equiv d^{2}\omega /dk^{2}$ of whistlers are given by 
\begin{widetext}
\begin{equation}
v_{g}=\left(  \frac{2c^{2}k}{\omega_{pe}^{2}}+\frac{g^{2}\hbar k}%
{4m_{e}\left(  \omega-\omega_{g}\right)  }\right)  /\left(  \frac{2\omega
}{\omega_{pe}^{2}}+\frac{\omega_c}{\left(  \omega-\omega_c\right)  ^{2}}%
+\frac{g^{2}\hbar k^{2}}{8m_{e}\left(  \omega-\omega_{g}\right)  ^{2}}\right)
, \label{14}%
\end{equation}%
\begin{equation}
v_{g}^{\prime}=\frac{v_{g}}{k}\left[  1-\frac{2kv_{g}^{2}}{\Lambda\omega
_{pe}^{2}}\left(  1-\frac{\omega_c\omega_{pe}^{2}}{\left(  \omega-\omega_c\right)
^{3}}\right)  -\frac{g^{2}\hbar k^{2}v_{g}}{4m_{e}\Lambda\left(  \omega
-\omega_{g}\right)  ^{2}}\left(  2-\frac{v_{g}k}{\omega-\omega_{g}}\right)
\right].
\end{equation}
\end{widetext}The nonlinear frequency shift $\Delta $ is given by%
\begin{equation}
\Delta =\frac{v_{g}}{\Lambda }\left[ \frac{k\omega v_{ez}}{\left( \omega
-\omega _{c}\right) ^{2}}+\left( \frac{\omega }{\omega -\omega _{c}}+\frac{%
g^{2}\hbar k^{2}}{4m_{e}\left( \omega -\omega _{g}\right) }\right) N\right] ,
\label{3}
\end{equation}%
where $\Lambda =2c^{2}k/\omega _{pe}^{2}+g^{2}\hbar k/4m_{e}\left( \omega
-\omega _{g}\right) \ $and $N\equiv n_{e}/n_{0}$ is the relative perturbed
density. By disregarding the spin contribution one can recover the previous
results \cite{WhistlerTheory1,WhistlerTheory2}. Note that the term $\propto
v_{ez},$ representing the Doppler shift due to the plasma streaming along
the external magnetic field, is no longer negligible, but may be comparable
to the other nonlinear terms, and can thus change the sign of the
nonlinearity as well. More precisely, both \ $v_{g}^{\prime }$ and\ $\Delta $
will change their sign depending on the frequency range to be considered as
well as the contribution from the spin correction terms. Later, we will see
that the change of sign is important for the formation of localized wave
packets at different whistler frequencies. The quantities $N$ and $v_{ez}$
are related to each other by the electron continuity equation.

\begin{equation}
\frac{\partial N}{\partial t}+\frac{\partial v_{ez}}{\partial z}=0.
\label{4}
\end{equation}%
Note that the ponderomotive force due to the EM whistlers usually drives the
lf (compared to the whistler wave frequency $\omega $) density perturbations
which propagate along the field lines with low-phase speed (compared to the
electron thermal speed). Thus, the lf electrostatic modulation also
satisfies the electron momentum equation 
\begin{align}
& \frac{\partial v_{ez}}{\partial t}+\frac{e}{m_{e}}E_{l}+V_{F}^{2}\frac{%
\partial N}{\partial z}-\frac{\hbar ^{2}}{4m_{e}^{2}}\frac{\partial ^{3}N}{%
\partial z^{3}}  \notag \\
& =\frac{e^{2}}{2m_{e}^{2}\omega ^{2}}\left( \Gamma _{1}\frac{\partial
|E|^{2}}{\partial z}-k\Gamma _{2}\frac{\partial |E|^{2}}{\partial t}\right) ,
\label{5}
\end{align}%
where $E_{l}$ is the lf part of the wave electric field and $V_{F}$ $=\sqrt{%
k_{B}T_{F}/m_{e}}$ \ is the Fermi speed relevant for a high density plasma 
\cite{FermiPressure}. Here $T_{F}\equiv \hbar ^{2}\left( 3\pi
^{2}n_{0}\right) ^{2/3}/2k_{B}m_{e}\ $and $k_{B}$ is the Boltzmann constant.
The term $\propto \hbar ^{2}$ is the quantum correction associated with the
Bohm de Broglie potential. The ponderomotive force contributions are
proportional to the constants $\Gamma _{1}$ and $\Gamma _{2}$ where%
\begin{align}
\Gamma _{1}& =\frac{\omega }{\omega -\omega _{c}}+\frac{g^{2}\hbar k^{2}}{%
4m_{e}\left( \omega -\omega _{g}\right) },  \notag \\
\Gamma _{2}& =\frac{\omega _{c}}{\left( \omega -\omega _{c}\right) ^{2}}+%
\frac{g^{2}\hbar k^{2}}{4m_{e}\left( \omega -\omega _{g}\right) ^{2}}.
\label{6}
\end{align}%
in which the first terms appear due to CPF \cite{WhistlerTheory2} and the
second ones $\left( \propto \hbar \right) $ are due to the SPF \cite%
{SpinPonderomotive}. The equations for the cold ion motion involved in the
lf IA perturbations are 
\begin{equation}
\frac{\partial n_{i}}{\partial t}+n_{0}\frac{\partial v_{iz}}{\partial z}=0,
\label{7}
\end{equation}%
\begin{equation}
\frac{\partial v_{iz}}{\partial t}=\frac{e}{m_{i}}E_{l},\text{ }  \label{8}
\end{equation}

\begin{equation}
\frac{\partial E_{l}}{\partial z}=\frac{e}{\varepsilon _{0}}(n_{i}-n_{e}).
\label{9}
\end{equation}%
Eliminating $n_{i},$ $E_{l},$ $v_{iz}$ and disregarding the term $\propto
m_{e}/m_{i},$ we obtain from Eqs. (\ref{5}), (\ref{7})-(\ref{9}) the driven
wave equation for lf perturbations of the Boussinesq-type as 
\begin{align}
& \frac{\partial ^{2}}{\partial t^{2}}\left( 1-\lambda _{F}^{2}\frac{%
\partial ^{2}}{\partial z^{2}}\right) N-c_{s}^{2}\frac{\partial ^{2}N}{%
\partial z^{2}}+\frac{\hbar ^{2}}{4m_{e}m_{i}}\frac{\partial ^{4}N}{\partial
z^{4}}  \notag \\
& =\mu _{1}\frac{\partial ^{2}}{\partial z^{2}}\left( \omega _{pi}^{2}+\frac{%
\partial ^{2}}{\partial t^{2}}\right) |E|^{2}-\mu _{2}\frac{\partial ^{2}}{%
\partial z\partial t}\left( \omega _{pi}^{2}+\frac{\partial ^{2}}{\partial
t^{2}}\right) |E|^{2},  \label{10}
\end{align}%
where $\mu _{1}=\varepsilon _{0}\Gamma _{1}/2n_{0}m_{e}\ $and $\mu
_{2}=\varepsilon _{0}k\Gamma _{2}/2n_{0}m_{e}\omega ^{2},$ $c_{s}=\sqrt{%
k_{B}T_{F}/m_{i}}$ is the ion-acoustic speed and $\lambda _{F}=c_{s}/\omega
_{pi}$ is the Fermi screening length for electrostatic oscillations.

Thus, we have a set of three coupled equations, namely (\ref{2}), (\ref{4})
and (\ref{10}), modified from previous results by the SPF and quantum
tunneling, which describes the nonlinear coupling of electron whistler waves
with the field aligned electrostatic density fluctuations. These equations
can be recast by normalizing the variables according to $z\rightarrow
z/\lambda _{F},t\rightarrow t\omega _{pi},E\rightarrow E/E_{0},$ $%
v_{ez}\rightarrow v_{ez}/c_{s}$, in which case we obtain%
\begin{equation}
i\left( \frac{\partial E}{\partial t}+V_{g}\frac{\partial E}{\partial z}%
\right) +\frac{V_{g}^{\prime }}{2}\frac{\partial ^{2}E}{\partial z^{2}}-\Psi
E=0,  \label{11}
\end{equation}%
\begin{equation}
\frac{\partial N}{\partial t}+\frac{\partial v_{ez}}{\partial z}=0,
\label{12}
\end{equation}%
and%
\begin{align}
& \left[ \frac{\partial ^{2}}{\partial t^{2}}\left( 1-\frac{\partial ^{2}}{%
\partial z^{2}}\right) -\frac{\partial ^{2}}{\partial z^{2}}+H^{2}\frac{%
\partial ^{4}}{\partial z^{4}}\right] N  \notag \\
& =\lambda _{1}\frac{\partial ^{2}}{\partial z^{2}}\left( 1+\frac{\partial
^{2}}{\partial t^{2}}\right) |E|^{2}-\lambda _{2}\frac{\partial ^{2}}{%
\partial z\partial t}\left( 1+\frac{\partial ^{2}}{\partial t^{2}}\right)
|E|^{2},  \label{13}
\end{align}%
where $E_{0}=\sqrt{2k_{B}T_{F}n_{0}/\varepsilon _{0}},$ $%
V_{g}=v_{g}/c_{s},V_{g}^{\prime }=v_{g}^{\prime }\omega _{pi}/c_{s}^{2},$ $%
\Psi =\Delta /\omega _{pi},$ $H=\hbar \omega _{pe}/2k_{B}T_{F}$ \ is the
quantum coupling parameter, $\lambda _{1}=\omega _{pe}^{2}\Gamma _{1}/\omega
^{2}\ $and $\lambda _{2}=\omega _{pe}^{2}kc_{s}\Gamma _{2}/\omega ^{2}.$
Equations (\ref{11})-(\ref{13}) contain the main results of the present
work. In particular, previous results \cite{WhistlerTheory1,WhistlerTheory2}
can be recovered by disregarding the spin contribution $\propto \hbar $ as
well as the particle dispersion $\propto H$ and considering, e.g., the
isothermal equation of state (relevant for low or moderate density plasmas).

\section{III. Stationary localized solutions}

In this section we will investigate the properties of nonlinear whistlers  by solving
numerically the Eqs. (\ref{11})-(\ref{13}) in
the stationary frame $\xi =z-Mt$ \ (where $M$ $\equiv V/c_{s})$. We will consider the parameter
regimes for the density and the magnetic field for which the nonrelativistic
fluid model is valid and SPF is comparable to the CPF. We will also see that
the case in which SPF dominates over the CPF may correspond to the strongly
magnetized superdense plasmas where relativistic treatement may be
necessary. However, before going further to such discussions let us first
consider the particular case in which the dispersion due to charge separtion
(quasineutrality) is negligibe. \ The latter can be justified even when the
spin effects dominate, i.e., $\chi \equiv \hbar k^{2}/m_{e}\omega \gg 1$ 
\cite{JPPMisra}. From the scaling%
\begin{equation}
k^{2}\lambda _{F}^{2}\sim \left( \frac{V_{F}^{2}}{c^{2}}\right) \left( \frac{%
c^{2}k^{2}}{\omega ^{2}}\right) \left( \frac{\omega ^{2}}{\omega _{pe}^{2}}%
\right) .  \label{14}
\end{equation}%
we find that\ the quasineutrality limit $k^{2}\lambda _{F}^{2}\ll 1$ holds
in nonrelativistic ($V_{F}^{2}\ll c^{2}$) plasmas as long as $n_{R}\equiv
ck/\omega >1\ $and $\omega _{pi}<\omega <$\ $\omega _{pe}$. \ However, we
will see that in a specific parameter regime, such restrictions can be valid
for very lf ($\omega \ll $\ $\omega _{c}$) whistler modes. In this case, $%
B_{0}<B_{Q}\equiv 4.4\times 10^{9}$ T and $n_{0}\gg 10^{32}\mathrm{m}^{-3}$ 
\cite{JPPMisra} with $T_{F}\gtrsim T_{B}\simeq \hbar \omega _{c}/k_{B}.$
Moreover, when $\chi \gg 1,$ the contribution from the term $\propto $ $%
\lambda _{2}$ can be smaller than that $\propto $ $\lambda _{1},$since $%
\left\vert \lambda _{2}/\lambda _{1}\right\vert \sim \left( kc_{s}/\omega
\right) \left( m_{e}\omega /\hbar k^{2}\right) \ll 1.$ Thus, in the
quasineutral regime, we obtain from Eqs. (\ref{11})-(\ref{13}) the following
NLS equation.%
\begin{equation}
\frac{V_{g}^{\prime }}{2}\frac{d^{2}E}{d\xi ^{2}}+i\left( V_{g}-M\right) 
\frac{dE}{d\xi }+\bar{\Delta}\left\vert E\right\vert ^{2}E=0,  \label{21}
\end{equation}%
together with \ 
\begin{equation}
N=\Lambda \left\vert E\right\vert ^{2},\text{ }v_{ez}=M\Lambda \left\vert
E\right\vert ^{2}.  \label{22}
\end{equation}%
Then we can write Eq. (\ref{3}) as $\Delta =-\bar{\Delta}\left\vert
E\right\vert ^{2}$, where $\bar{\Delta}$ is defined as 
\begin{equation}
\bar{\Delta}\approx -\frac{\Lambda \omega v_{g}}{\Lambda \omega _{pi}\left(
\omega -\omega _{c}\right) }\left( 1+\frac{kV}{\omega -\omega _{c}}+\frac{%
\hbar k^{2}}{m_{e}\omega }\right) ,
\end{equation}%
where $\Lambda =\left( \lambda _{1}+\lambda _{2}M\right) /\left(
M^{2}-1\right) .$ Physically, the electrons experience a longitudinal force
exerted by the front of the whistler pulse, and thereby gain a net energy.
The electrons gain energy during the rising front of the pulse, but then
slows down by the backward ponderomotive-like force. Moreover, electrons can
approach the group velocity of the whistler when it reaches the pulse peak
at the center. From Eq. (\ref{22}), we find that this can be possible for $%
M^{2}\ll 1$, which may happen for a whistler frequency satisfying $\omega
_{c}/2<\omega <\omega _{c}$ and for high density ($\sim 10^{36}$m$^{-3}$)
and strongly magnetized ($B_{0}\sim 10^{8}$T) plasmas. In this case, the
Fermi speed may exceed the group speed ($\sim c$). On the other hand,
corresponding to the parameters as in Fig. 1 below, $M\gg 1$ and $\lambda
_{1}+\lambda _{2}M\sim 0.01,$ so that $N\sim \left\vert E\right\vert
^{2}\times 10^{-7},$ $v_{ez}\sim \left\vert E\right\vert ^{2}\times 10^{-5}$
and $\bar{\Delta}\sim 10^{-7}.$ Again, note that slow electrons can freely
move along the direction of the external magnetic field. The finite velocity
perturbations would then induce an additional density change in order to
maintain the conservation of particles (equation of continuity) under
localized disturbances. Consequently, the total density variation in the
frequency-shift becomes $\delta N,$ where

\begin{equation}
\delta \approx 1+\frac{kV}{\omega -\omega _{c}}+\frac{\hbar k^{2}}{%
m_{e}\omega }  \label{delta}
\end{equation}%
\begin{figure}[tbp]
\begin{center}
\includegraphics[height=3in,width=3.0in]{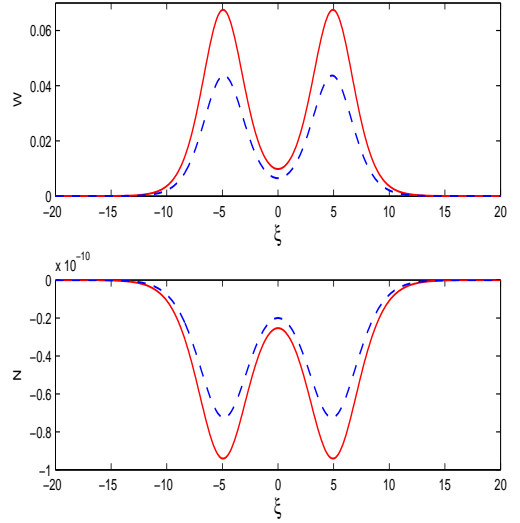}
\end{center}
\caption{(Color online) Whistler solitary solution of Eqs. (\protect\ref{18}%
) and (\protect\ref{19}) with associated electric field $W$ (upper panel)
and density perturbation $N$ (lower panel) for $\Theta =0.8$ (solid line)
and $0.82$ (dashed line). The other parameter values are $n_{0}=10^{34}$m$%
^{-3}$, $B_{0}=5\times 10^{8}$T, $\protect\omega _{c}/\protect\omega %
_{pe}=15.6$, $\protect\omega =0.4,$ $M(\equiv
v_{g}/c_{s})=234.54,V_{g}^{\prime }=-1.81,v_{g}=2.98\times 10^{8}$m/s$.$ }
\end{figure}
\begin{figure}[tbp]
\begin{center}
\includegraphics[height=3in,width=3.0in]{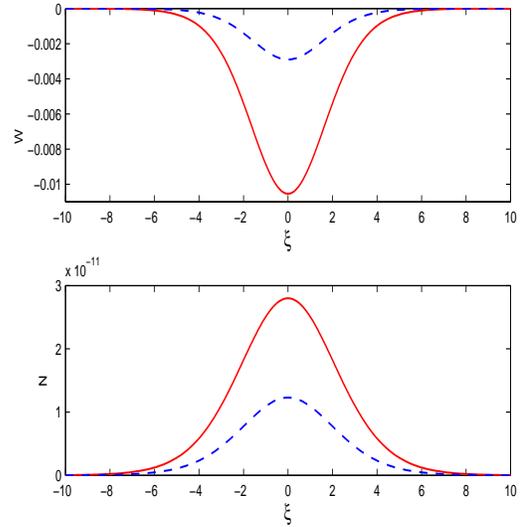}
\end{center}
\caption{(Color online) Whistler solitary solution of Eqs. (\protect\ref{18}%
) and (\protect\ref{19}) with associated electric field $W$ (upper panel)
and density perturbation $N$ (lower panel) for $B_{0}=5.8\times 10^{8}$T
(solid line) and $5.9\times 10^{8}$T (dashed line). The other parameter
values are $n_{0}=10^{34}$m$^{-3}$, $\Theta =0.8$, $\protect\omega =0.4.$
The corresponding $M$ values are $M=234.88$ (solid line) and $234.91$
(dashed line). }
\end{figure}
Clearly, $\delta $ changes sign whenever the third term $\propto \hbar $ \
in Eq. (\ref{delta}) dominates over the other two terms. Now, for lf
propagation of whistlers, $\lambda _{1}<0\ $and as in the previous section, $%
\lambda _{2}M$ $(>0)$ is smaller compared to $\lambda _{1}$ when the spin
contribution dominates. Thus, in the quasineutral lf regime, the density and
velocity perturbations are positive and negative according as the whistler
wave propagation is subsonic or supersonic [see Eq. (\ref{22})].

Furthermore, localized bright (dark) envelope solutions of Eq. (\ref{21})
exist through the modulational instability (stability) when $\bar{\Delta}%
v_{g}^{\prime }>0(<0).$ For lf waves $(\omega <\omega _{c}),$ when $\hbar
k^{2}/m_{e}\omega \gg 1$, $\lambda _{1}+\lambda _{2}M<0,$ $v_{g}^{\prime }<0 
$ and\ $\bar{\Delta}\gtrless 0$ according as $M\lessgtr 1.$ Hence,\ a
possible final state of the MI could be a supersonic (subsonic) bright
(dark) soliton-like structure in a quasineutral spin quantum plasma. \
Equation (\ref{21}) has an exact soliton solution (when $\bar{\Delta}$ and $%
V_{g}^{\prime }$ have the same sign) of the form 
\begin{equation}
E(\xi )=E_{m}\text{sech}\left[ E_{m}\sqrt{\bar{\Delta}/V_{g}^{\prime }}%
\left( \xi -\xi _{0}\right) \right] ,
\end{equation}%
where $E_{m},\xi _{0}$ are constants. The other particular cases, namely the
quasistationary\ lf density response (i.e., $\partial _{t}\rightarrow 0)$ 
for which $\omega<<\omega_{pi}$ \cite{WhistlerTheory1} and the case of unidirectional propagation (near
sonic envelope) in which the quasineutrality is not a valid assumption \cite{WhistlerTheory3(nearsonic)} will not be discussed here
as those cases are not so relevant to the parameter regimes to be
considered, instead we will focus on our main Eqs. (\ref{11})- (\ref{13}).

Thus, we look for stationary solutions of\ Eqs. (\ref{11})- (\ref{13}) in
the stationary frame $\xi =z-Mt$.\ Here we assume $E$ to be of the form $%
E=W\left( \xi \right) \exp \left( -i\Theta t\right) ,$where $W$ is a real
function and $\Theta $ is a real constant. Then Eqs. (\ref{11})- (\ref{13})
reduce to 
\begin{equation}
\frac{V_{g}^{\prime }}{2}\frac{d^{2}W}{d\xi ^{2}}+W\Omega +\tilde{\Delta}%
NW=0,  \label{18}
\end{equation}%
\begin{align}
& \left( -M^{2}+H^{2}\right) \frac{d^{2}N}{d\xi ^{2}}+\left( M^{2}-1\right) N
\notag \\
& =\left( \lambda _{1}+\lambda _{2}M\right) \left( M^{2}\frac{d^{2}W^{2}}{%
d\xi ^{2}}+W^{2}\right) ,  \label{19}
\end{align}%
where $\tilde{\Delta}=\bar{\Delta}/\Lambda .$ We numerically solve the
equations (\ref{18}) and (\ref{19}) by Newton method with the boundary
conditions $N,$ $W,$ $d_{\xi }^{2}N,$ $d_{\xi }^{2}W\rightarrow 0$ as $%
\left\vert \xi \right\vert \rightarrow \infty $. We consider the density and
magnetic field strength to vary as $n_{0}\sim 10^{34}-10^{36}$m$^{-3}$ and $%
B_{0}\sim 10^{8}$T. Figure 1 illustrates the existence of double-hump
localized whistler envelope accompanied with a density depletion for a set
of parameters: $n_{0}\sim 10^{34}$m$^{-3},$ $B_{0}\sim 5\times 10^{8}$T, $%
\omega =0.4\ $\ and $\Theta =0.2.$ The corresponding frequencies are $\omega
_{pi}=1.32\times 10^{17}$s$^{-1},$ $\omega _{pe}=5.64\times 10^{18}s^{-1}$
and $\omega _{c}=8.79\times 10^{19}$s$^{-1}.$ Also, $M(\equiv
v_{g}/c_{s})=234.54,$ $\lambda _{De}(=\lambda _{Fe}\equiv V_{Fe}/\omega
_{pe})=9.67\times 10^{-12}$m and $V_{Fe}=5.46\times 10^{7}$m/s. Thus, the
whistlers have negative group dispersion with $V_{g}^{\prime }=-1.81.$ From
the dispersion relation we obtain $k=1.18\times 10^{11}$m$^{-1},$ which
corresponds to whistlers with a wavelength of $5.3121\times 10^{-11}$m, and
the group speed is $v_{g}=2.98\times 10^{8}$m/s. Furthermore, the nonlinear
frequency shift is obtained as $\tilde{\Delta}=0.85$. The density depletion
is observed quite small due to large group velocity (compared to the sound
speed) of the whistler waves.

\ In another illustration (Fig. 2) with a higher magnetic field, we observe
a dark-soliton-like structure correlated with a density hump. The amplitude
of the solitary pulse decreases as the magnetic field increases. \ In Fig. 3
we have presented the solitary structures when the density is very high ($%
n_{0}\sim 10^{36}$m$^{-3}$). This basically corresponds to the case when\ $%
\hbar k^{2}/m_{e}\omega \gtrsim 1.$ However, in this case one must note that
the Fermi speed is close to or can even be larger than the speed of light in
vacuum and so, nonrelativistic quantum fluid model may no longer be
appropriate. The quantum parameter $H$ has no significant role for the
regime considered here, as can be seen that $M(\equiv V_{g})$ mainly
dominates in the term $-M^{2}+H^{2}$ [Eq. (\ref{19})], because of large
group velocity ($\approx c$). In order that $H$ can be comparable to $M$,
one might have to consider relatively higher densities $(>10^{36}$m$^{-3})$
and weakly magnetized $(\ll 10^{8}T)$ plasmas. However, in this case the
coefficient $\lambda _{1}+\lambda _{2}M$ ($\sim 10^{5}$) will be much larger
than the other coefficients, which might prevent any hope for localized
solution. As shown in Fig. 4, one can excite a nondiverging whistler with a
positive group dispersion in other regime, e.g., $\omega =0.189,$ $\Theta
=0.7,$ $n_{0}\sim 7\times 10^{36}$m$^{-3}$ and $B_{0}\sim 5\times 10^{8}$T
for which $V_{g}^{\prime }=0.262,$ $H=0.11,$ $v_{g}=2.37\times 10^{8}$m/s, $%
M=45.24$, $V_{Fe}=2.25\times 10^{8}$m/s, $c_{s}=5.24\times 10^{6}$m/s. This
basically corresponds to oscillatory pulse associated with a field-aligned
density hump ($N\sim 10^{-10}).$ 
\begin{figure}[tbp]
\begin{center}
\includegraphics[height=3in,width=3.0in]{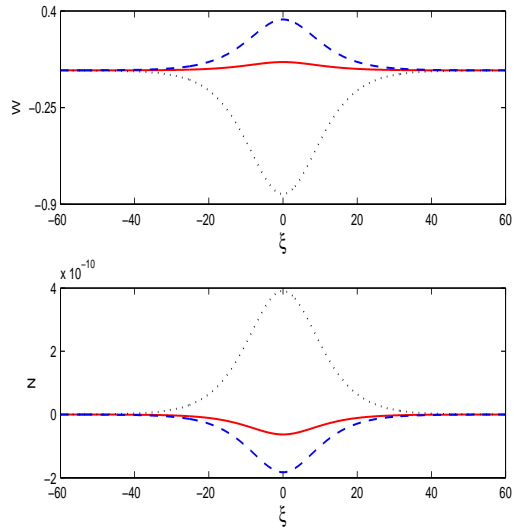}
\end{center}
\caption{(Color online) Whistler solitary solution of Eqs. (\protect\ref{18}%
) and (\protect\ref{19}) with associated electric field $W$ (upper panel)
and density perturbation $N$ (lower panel) for $\protect\omega =0.4$ (solid
line) and $0.38$ (dashed and dotted ine). The other parameter values are $%
B_{0}=5\times 10^{8}$T, $n_{0}=10^{36}$m$^{-3}$ (for solid and\ dashed line)
and $2.1\times 10^{36}$m$^{-3}$ (for dotted line), $\Theta =0.2$. The values
of $M$ are $M=35.38$ (for solid and dashed line) and $22.23$ (for dotted
line). }
\end{figure}
\begin{figure}[tbp]
\begin{center}
\includegraphics[height=3in,width=3.0in]{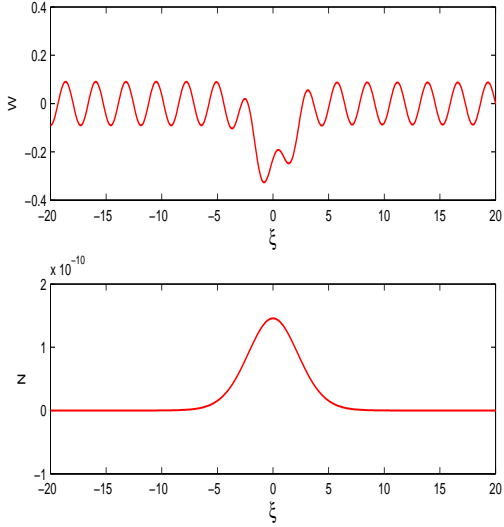}
\end{center}
\caption{(Color online) Ducted whistler obtained as solution of Eqs. (%
\protect\ref{18}) and (\protect\ref{19}) with associated electric field $W$
(upper panel) and density perturbation $N$ (lower panel) for $\protect\omega %
=0.189,$ $B_{0}=5\times 10^{8}$T, $n_{0}=7\times 10^{35}$m$^{-3}$, $\Theta
=0.7$. The other parameters are $M=45.24,$ $V_{g}^{\prime
}=0.262,H=0.11,v_{g}=2.37\times 10^{8}$m/s. }
\end{figure}

\section{IV. Growth rate of instability}

Nonlinear interaction of the hf \ pump EM whistlers $(\omega,k)$ with lf
electrostatic field aligned perturbations $(\Omega,K)$ gives rise upper and
lower side bands with frequency and wave numbers respectively $(\omega
+\Omega,k+K)$ and $(\omega-\Omega,k-K).$ The latter interacts with the pump
and thus produces a lf ponderomotive force which eventually reinforces the
lf electrostatic oscillations. When all the perturbations are aligned along
the external magnetic field, the parametric interactions of EM waves can be
described from Eqs. (\ref{11})- (\ref{13}) by the following dispersion
relation.

\begin{align}
& \Lambda K^{2}V_{g}^{\prime }\left[ K^{2}\left( 1+H^{2}K^{2}\right) -\left(
1+K^{2}\right) \Omega ^{2}\right]  \notag \\
& =4V_{g}E_{0}^{2}\left( 1-\Omega ^{2}\right) \left( \lambda _{1}K+\lambda
_{2}\Omega \right) \left( K\Gamma _{1}+\zeta \Omega \right) ,  \label{23}
\end{align}%
where $\zeta =k\omega \omega _{c}^{3}/\lambda _{De}\left( \omega -1\right)
^{2}$ and $\omega ,$ $k$ have been normalized by $\omega _{c}$ and $\lambda
_{De}^{-1}$ respectively. Some simplification can be in order. Note that
under the quasineutrality assumption, the coefficient of $\Omega ^{4},$ $%
\Omega ^{3},$ and the term $\propto K^{4}\ $as well as the term $\propto
\lambda _{1}$ in the coefficient of $\Omega ^{2}$ will not appear. Also, for
lf propagation of whistlers $\left( \omega <\omega _{c}\right) ,$ $\lambda
_{2}$ is smaller and thus being neglected. Moreover, the ratio of the term $%
\propto \zeta $ in the coefficient of $\Omega $ (which appears due to the
parallel electron streaming $v_{ez}$ ) and the constant term $\propto \Gamma
_{1}$ scales as $\left( k/K\right) \omega \left( m_{e}\Omega \omega
_{c}/\hbar k^{2}\right) $ and we need $\left( k/K\right) \omega \left(
m_{e}\Omega \omega _{c}/\hbar k^{2}\right) \ll 1$ for spin effects to be
dominant. Thus, in this case the dispersion relation reduces to 
\begin{equation}
\Omega ^{2}\approx K^{2}\left( 1+H^{2}K^{2}\right) -E_{0}^{2}\eta ^{2},
\end{equation}%
where $\eta ^{2}=\pm 4V_{g}\lambda _{1}\Gamma _{1}/\omega _{c}^{2}\Lambda
V_{g}^{\prime }$ in which $\omega ,k$ etc. are being normalized. Clearly,
MI\ sets in for modulation wave numbers satisfying $K\sqrt{1+H^{2}K^{2}}%
<E_{0}\eta ,$ or $K<K_{c}\approx E_{0}\eta $ for highly dense medium and
small $K.$ The growth rate of instability $(\Omega =i\gamma )$ is then given
by

\begin{equation}
\gamma\approx\sqrt{E_{0}^{2}\eta^{2}-K^{2}\left( 1+H^{2}K^{2}\right) }.
\end{equation}
Hence, in the long-wavelength limit ($K\rightarrow0)$ maximum growth rate of
instability can be achieved, and is roughly proportional to the pump wave
electric field $E_{0}$ and $\eta.$ For parameters as in Fig. 1, we obtain $%
\gamma\approx2.77.$ It basically restricts the characteristic length-scale
to a certain value for the formation of envelope solitons through MI.

\section{V. Discussion and Conclusion}

In the present investigation focusing on whistler waves we point out that
the spin contribution is substantial when $\hbar k^{2}/m_{e}\omega \gg 1,$
i.e., when $\hbar \omega _{c}/m_{e}c^{2}>1$ and $\omega ^{2}\ll c^{2}k^{2}.$
This corresponds to the case in which the magnetic field strength, $%
B_{0}\gtrsim B_{Q}$ and the particle density is very high, i.e., $%
n_{0}\gtrsim 10^{36}$m$^{-3}$ for which the magnetic field is non-quantizing
and does not affect the thermodynamic properties of electrons. However, in
such regimes, the Fermi velocity may approach or exceed the whistler group
velocity (close to $c$ in the present study), and so the nonrelativistic
quantum fluid model may no longer be appropriate to consider. In the present
work, we have considered $B_{0}\sim 10^{8}$T and the density to vary in the
range $10^{34}\lesssim n_{0}\lesssim 10^{36}$m$^{-3}$ in order that the
nonrelativistic fluid model is valid to some extent. Moreover that $\omega
_{pe}\gtrsim \omega _{c}$ and the terms due to spin magnetization current
together with the SPF are comparable to the classical counter parts.
Furthermore, in this regime the velocity of electrons remains much smaller
than the whistler group velocity ($\sim c$).

Since the whistler group speed is much higher than the IA speed, whistler
solitons are not significantly affected by the particle dispersion
associated with the Bohm potential as well as the Fermi-Dirac pressure,
though \ the length scale of excitation is of the order of the Compton
wavelength. However, those effects reduce the plasma characteristic
wavelength of excitation. \ Such effects can be more significant in some
other regimes when $M\lesssim 1$ and/ or for possible excitation of the ion
wakefields at nanoscales. Note that since degenerate electrons follow the
Fermi-Dirac pressure law (where the Fermi temperature is density dependent), the
cold plasma limit can not be recovered from the present study unless
one consideres, e.g., isothermal equation of state to be relevant for low or
moderate density plasmas. Furthermore, $H\rightarrow 0$ means that one
approaches the higher density regimes and $H=0$ is the case when one simply
disregards the quantum tunneling effect. 

The parameter regimes considered
here can be achievable in the magnetized white dwarfs ($\sim 10^{36}$m$^{-3}$%
) as well as in the next generation intense laser-solid density plasma
experiments ($\sim 10^{34}$m$^{-3}$), in x-ray free electron lasers, and in
plasmonic devices. One can, in principle, go beyond the parameter regimes
considered here (since there is no specific theoretical limit for the
density), however, we have to be careful about those parameter values for
the excitation of localized whistlers and for spin-ponderomotive force to
have a role. The latter may dominantly accelerate the ions by separating the
electric charges and building up a high electric field. However, plasma can
sustain such high electric fields, and so it remains an attractive medium
for particle acceleration, which is still a most important areas of research
works in both laboratory and astrophysical plasmas.

In conclusion, we have presented a new set of nonlinear equations which
governs the dynamics of modulated whistlers interacting with\ the
field-aligned electrostatic lf density perturbations due to IA fluctuation,
in a magnetized spin quantum plasma. Both the classical as well as the
spin-induced ponderomotive force has been considered to modify the local
plasma density in a self-consistent manner. Numerical simulation of the
governing equations in the stationary frame [Eqs. (\ref{18}) and (\ref{19})]
reveals the existence of supersonic stationary envelope solitons
characterized by a single or double hump whistler wave electric fields that\
are trapped in a self-created density cavity. This happens for wave
frequency satisfying $\omega<\omega_{c}/2$ and when the whistler has
negative group dispersion. When the whistler frequency is smaller than $%
\omega_{c}/4$ and the group dispersion is positive at higher densities, one
can excite a nondiverging whistler wave, i.e., a ducted whistler.\ The
latter corresponds to a field aligned density hump with $N\sim10^{-10}.$
Furthermore, the whistler solitons with density dips and humps can occur
depending on the consideration of the frequency regime as well as the
magnetic field strength and/ or the particle density. 

We ought to mention that our present investigation on the nonlinear propagation of EM whistlers
might play an important role in studies of beat-wave particle accelerators 
\cite{BeatWave} as well as in the problem of radio-frequency
electron-cyclotron-resonance heating \cite{ResonanceHeating} of plasmas
where the driver, instead of being a laser, is a whistler wave.

\textbf{Acknowledgment}

APM is grateful to the Kempe Foundations, Sweden, for support through
Grant No. SMK-2647. MM was supported by the European Research Council under Contract
No. 204059-QPQV and the Swedish Research Council under Contract No. 2007-4422.

\end{document}